\documentclass{article}
\usepackage{spconf,amsmath,graphicx}
\usepackage{array}
\usepackage{amsmath}
\usepackage{amsfonts}

\newcolumntype{P}[1]{>{\centering\arraybackslash}m{#1}}
\title{Deep Clustering for General-Purpose Audio Representations}
\name{ Sreyan Ghosh$^{\star \dagger}$ \qquad Sandesh V Katta$^{\star \dagger}$ \qquad  Ashish Seth$^{\dagger}$ \qquad S. Umesh$^{\dagger}$ \thanks{\hspace*{-1mm}$^{\star}$These authors contributed equally to this work}}
\address{$^{\dagger}$ Speech Lab, Dept. of Electrical Engineering, IIT Madras, Chennai, India \\}

\begin{document}
%
\maketitle
\begin{abstract}
We introduce \textbf{DECAR}, a self-supervised pre-training approach for learning general-purpose audio representations. Our system is based on clustering: it utilizes an offline clustering step to provide target labels that act as pseudo-labels for solving a prediction task. We develop on top of recent advances in self-supervised learning for computer vision and design a lightweight, easy-to-use self-supervised pre-training scheme. We pre-train DECAR embeddings on a balanced subset of the large-scale Audioset dataset and transfer those representations to 9 downstream classification tasks, including speech, music, animal sounds, and acoustic scenes. Furthermore, we conduct ablation studies identifying key design choices and also make all our code and pre-trained models publicly available
\footnote[1]{https://github.com/Speech-Lab-IITM/DECAR}.
\end{abstract}
\section{Introduction}
\label{sec:intro}

Self-supervised representation learning aims at obtaining features without using manual annotations, and is rapidly closing the performance gap with supervised pre-training in a wide range of tasks across Computer Vision (CV) \cite{caron2019deep,chen2020simple}, Natural Language Processing (NLP) \cite{devlin2019bert}, and Speech Processing, especially Automatic Speech Recognition (ASR) \cite{baevski2020wav2vec,hsu2021hubert}. These systems leverage self-supervised learning algorithms to learn representations by solving various tasks on large, unlabeled datasets, and the weights learned are then either transferred for solving other downstream tasks or used as a feature extractor for input into a system, for learning a different set of weights for another task.

The goal of speech and audio representation learning is to 
learn a transformation from the acoustic signal that makes high-level information more accessible to downstream tasks. In the acoustics domain, the recent success of unsupervised speech representation learning
has shown to outperform low-level features like mel-spectrograms,  waveform, or filter-banks, in various Spoken Language Processing (SLP) downstream tasks, the primary reason being that the former tends to learn more detailed information from speech. Despite recent progress, most work on self-supervised learning for speech and audio processing focus mostly on ASR and ignores other speech and audio tasks such as acoustic scene detection or animal vocalizations, except \cite{jansen2017unsupervised,tagliasacchi2019selfsupervised,saeed2020contrastive,Shor_2020}. Triplet-based objectives used in \cite{jansen2017unsupervised,Shor_2020}, heavily rely on the mining of negative samples, and the quality of learned features varies significantly with the sample generation scheme  \cite{tagliasacchi2019selfsupervised}. The methodology used in  \cite{saeed2020contrastive} learns general-purpose audio representations by solving a contrastive task. Contrastive learning systems typically work online and rely on a large number of explicit pairwise feature comparisons, which is computationally challenging, given the fact they require large batches for mining negative samples.

In this paper, we introduce \textbf{DECAR}: \textbf{DE}ep \textbf{C}lustering for General-Purpose \textbf{A}udio \textbf{R}epresentations, a simple self-supervised pre-training framework to learn general-purpose audio representations of sounds beyond and including speech. Our system requires no prior knowledge, minimal additional steps, alleviates dependency on large batches, and closely resembles a general supervised training procedure. The DECAR framework employs an offline clustering step to generate noisy labels from discrete audio samples and learns by predicting cluster assignments. We show that it is possible to obtain useful general-purpose audio features with a clustering framework. 


Our DECAR pre-training approach consists of two forward passes of the entire dataset for each epoch. At each epoch, first, we cluster the output of our ConvNet feature extractor for all audio samples, and then use the subsequent cluster assignments as “pseudo-labels” for an augmented versions of the same audio samples to optimize the network. Formally, our approach iteratively learns features and groups them.


We demonstrate the effectiveness of DECAR over a range of challenging and diverse downstream tasks, including speech, music, acoustic scenes, and animal sounds, and intent to highlight its performance on low resource downstream tasks. After pre-training on a relatively smaller subset of the large-scale AudioSet database \cite{7952261}, compared to prior work, we show that only a linear classifier trained over DECAR embeddings is competitive in performance to fully-supervised in-domain convolutional network's performance and exceeds it when fine-tuned end-to-end, with significant performance boosts in low-resource downstream tasks. These experiments demonstrate that DECAR offers a simple, easy-to-implement unsupervised audio pre-training methodology for large-scale end-to-end training and learns audio representations that can be generalized across a variety of audio tasks, including speech, songs, musical instruments, etc. 


\section{Related Work}
The past decade has seen impressive success in self-supervised learning in vision, speech, and text, pushing boundaries in low resource settings and improving performance in various downstream tasks. Methodologies to learn representations from discrete or continuous input sequences, such as in Speech or Natural Language Processing applications, use either masked modeling, by solving different tasks like contrastive learning \cite{baevski2020wav2vec}, reconstruction \cite{Liu_2020} and class prediction \cite{devlin2019bert,hsu2021hubert}, or auto-regressive generation of input sequences by encoding temporal information of past acoustic sequence \cite{chung2019unsupervised}.\\ 
\begin{figure*}[tp]
  \centering
  \includegraphics[width=1\textwidth]{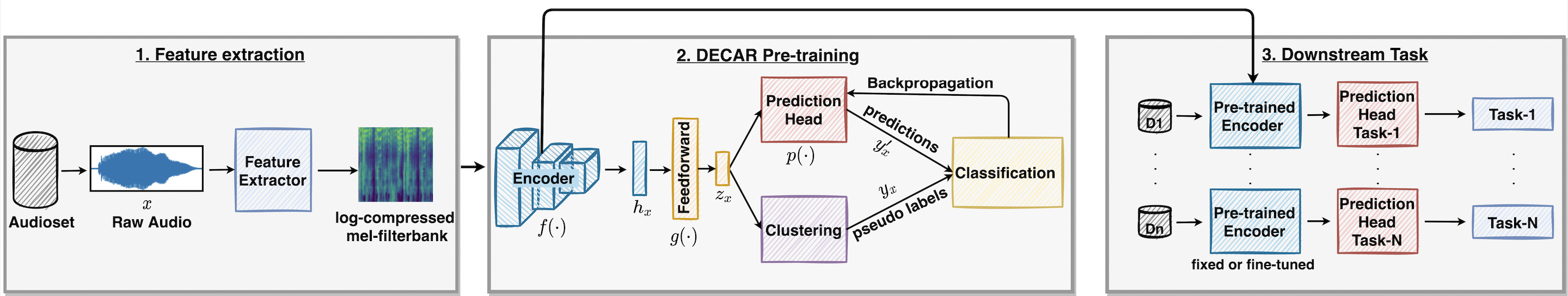}
  \caption{Pretraining and Finetuning of DECAR}
  \label{fig:lab1}
  \vspace{\floatsep}
\end{figure*}

In recent times, contrastive learning has shown great success in various settings like masked modeling in speech \cite{baevski2020wav2vec}, or in vision where the algorithm enforces similarity between pairs of different augmentations for the same image \cite{chen2020simple}. However, the bottleneck of requiring large batches to mine distractors has led to the recent proposal of clustering-based methodologies, which learns representations by predicting pseudo-labels assigned to instances in the form of cluster assignments, both in speech, \cite{hsu2021hubert,maekaku2021speech} and vision domain  \cite{caron2019deep,caron2021unsupervised}. While HuBERT \cite{hsu2021hubert} exploits clustering to solve a masked prediction problem on speech frames in an End-to-End transformer framework, for self-supervised
\emph{speech} representation learning, \cite{maekaku2021speech} uses it for solving a different task in a multi-task learning framework. To the best of our knowledge, there is no prior work in this domain that exploits a clustering framework for self-supervised audio representation learning. 

\section{Methodology}
\label{sec:method}

We learn general-purpose audio representations from unlabeled data by pre-training a neural network with a clustering approach. Our approach, inspired from the DeepCluster framework \cite{caron2019deep}, is based on pre-training a convolutional feature extractor on unlabeled audio data by predicting pseudo-labels which are obtained by clustering vector embedding representations \emph{e} of audio samples from a large-scale dataset and later combining our feature extractor with an additional classification layer for solving various audio understanding tasks across several downstream tasks.

As discussed earlier, our pre-training step makes two passes through our entire dataset at each epoch. At the first forward pass for each epoch, for each audio sample $x$, we extract the  log-compressed mel-filterbanks and pass it through our ConvNet feature extractor $f(.)$ to obtain a single embedding $ f(x) = h_x \in \mathbb{R}^{n}$. Then we pass these embeddings through a feedforward layer $g(.)$, to obtain $ g(h_x) = z_x\in \mathbb{R}^{k}$. Post this step, the outputs $z_x$ 
are clustered using our clustering algorithm. Before clustering, $z_x$ is PCA-reduced to 128 dimensions, whitened, and l2-normalized. We experiment with both K-Means and Power Iteration Clustering (PIC) \cite{lin2010power} as choices for our clustering algorithm.


\begin{equation}
\label{eqn:cluster}
  \min _{C \in \mathbb{R}^{d \times k}} \frac{1}{N} \sum_{x=1}^{N}  \min _{y_{x} \in\{0,1\}^{k}}\left\|z_{x} -C y_{x}\right\|_{2}^{2}  \text{ } \ni \text{ } y_{n}^{\top} 1_{k}=1
\end{equation}

At the second pass, we predict the cluster assignment for each audio sample assigned during the first step with an augmented version of the same sample. To achieve this, we add a \emph{prediction-head} which is a linear layer and learns a parameterized function $p(.)$ where $ p(z_x) = y^{\prime}_x \in \mathbb{R}^{c}$ and \emph{c} is the number of cluster centroids. Finally, we pass $y^{\prime}_x$ through a softmax activation function to get the probability distribution of each cluster class for each audio sample and train all the parameters of the network by minimizing Cross-Entropy loss over the real and predicted class distribution (\ref{eqn:cross_entropy}).

\begin{equation}
\label{eqn:cross_entropy}
\text {C.E.}=-\frac{1}{N} \sum_{x=1}^{N} \sum_{j=1}^{M} y_{x_j} \log \left(\operatorname{softmax}(y^{\prime}_{x})_{_j}\right)
\end{equation}





Formally, \textbf{DECAR} alternates between clustering the features to produce pseudo-labels by solving (\ref{eqn:cluster}), where it jointly learns a ${d \times k}$ centroid matrix $C$ and cluster assignments $y_x$, and updating the parameters of the ConvNet by predicting these pseudo-labels using (\ref{eqn:cross_entropy}). Additionally, during the second step, while assigning the pseudo-labels to the audio samples post clustering, we perform SpecAugment augmentation \cite{Park_2019} on the extracted log-compressed Mel-filterbanks before passing it on to the ConvNet for the prediction task when training the network. We hypothesize that this improves the robustness of our learned weights and also improves performance by enforcing invariance to data augmentation.


We handle the problem of trivial parameterization during the cluster label prediction step, where a vast majority of all the audio samples in a batch might belong to the same cluster, and minimizing (\ref{eqn:cross_entropy}) might lead to a trivial solution, by using a custom sampler, which samples batches of audios based on a uniform distribution over all the pseudo-labels.

\section{Datasets}
\label{sec:dataset}
We pre-train DECAR embeddings on a balanced subset of the diverse, large-scale Audioset \cite{7952261}. The complete dataset contains 2 million excerpts of 10 seconds of audio from YouTube videos that are annotated in a multi-label
fashion with 527 classes. Our subset consists of 10\% of the total number of audio samples from the original dataset sampled. Before our work, this dataset has also been used \cite{tagliasacchi2019selfsupervised,saeed2020contrastive,Shor_2020} for self-supervised pre-training. Since our pre-training strategy is self-supervised, we don't utilize the labels in the train set.

We perform the downstream evaluation on a variety of tasks, including both speech and non-speech tasks. For the selection of datasets for these tasks, we are inspired by previous work in the self-supervised audio pre-training domain \cite{tagliasacchi2019selfsupervised,saeed2020contrastive,Shor_2020}, with some of our additions to take into account more low-resource tasks. For speaker identification, we use two datasets, namely, LibriSpeech \cite{7178964} and Voxceleb \cite{Nagrani_2017}. For keyword spotting, we use Google Speech Commands V1, and V2 datasets \cite{warden2018speech}. For acoustic scene classification, we use TUT Urban Acoustic Scenes 2018 (TUT), consisting of labeled audio segments from 10 different acoustic scenes. For animal vocalizations, we use the Bird Song Detection (BSD) dataset \cite{stowell2019automatic} to solve a binary classification problem \footnote{The original test labels are no more available, and thus we took a random split of 80:20 for train and test from the combined development sets}. For Speech Emotion Recognition (SER) we use IEMOCAP dataset \cite{busso2008iemocap}. For music recognition, we use the NSynth dataset \cite{engel2017neural} of musical notes from different instruments. Finally, for language identification, we use the Voxforge dataset  categorize audio clips of different classes of spoken language.

More details about the downstream datasets can be found in Table \ref{downstream_Dataset_summary}.

\begin {table*}[t]
\small
\caption{Dataset statistics for downstream benchmark tasks.}
\vspace{1mm}
\begin{tabular}{P{0.25\textwidth}@{}P{0.23\textwidth} P{0.13\textwidth}@{}P{0.13\textwidth}@{}P{0.15\textwidth}}
\hline
\textbf{Dataset} & \textbf{Target} & \textbf{No of  classes}  & \textbf{No of  samples} & \textbf{Avg  duration (sec)}    \\
\hline 
Libri Speech  & Speaker Identification &  585 & 28538 & 12.69 \\
Voxceleb 1  & Speaker Identification &   1211 & 153397 & 8.20 \\
IEMOCAP  & Emotion recognition &  4 & 4490 & 4.49\\
TUT Urban 2018 (TUT) & Acuostic scene classification & 10 & 8640 & 10.00 \\
 Speech Commands V1 & Key word recognition &   12 & 64721 & 0.98 \\
 Speech Commands V2 & Key word recognition &   12 &  105829 &0.98\\
Bird Song Detection(BSD) & Song detection   & 2 & 15690 &10.08\\
Voxforge & Language Identification & 6 & 8803 &6.68\\
N-synth & Music Identification & 11 & 301883  &4.00\\
\hline
\end{tabular}
\label{downstream_Dataset_summary}
\end{table*}

\section{Experimental Setup}

Given an audio input sequence, we extract log-compressed Mel-filterbanks with a window size of 25 ms, a hop size of 10 ms, and N = 64 Mel-spaced frequency bins in the range 60–7800 Hz. Our ConvNet feature encoder $h$ is based on the lightweight and highly scalable EfficientNet-B0. We apply global max-pooling to the last layer of the encoder to get an embedding $h_x$ of size 1280. During pre-training, we pass $h_x$ through a linear layer $g$, which contains a fully connected layer with 512 units followed by the Relu activation function and a prediction head $p$ (for the 2\textsuperscript{nd} pass) with units equal to the number of cluster centroids. We pre-train for a total of 100 epochs with a patience of 10 epochs with the criterion of no improvement in the Normalized Mutual Information (NMI) between the cluster assignments of the current epoch $t$ and the previous epoch $t-1$, which are denoted as $A$ and $B$ in (\ref{eqn:nmi}) respectively. We learn the parameters of our model with Stochastic Gradient Descent (SGD) and a learning rate of 0.05 in batched-mode with a batch size of 64.

\begin{equation}
\label{eqn:nmi}
\operatorname{NMI}(A ; B)=\frac{\mathrm{I}(A ; B)}{\sqrt{\mathrm{H}(A) \mathrm{H}(B)}}
\end{equation}

Post-self-supervised pre-training, we discard the linear layer $g(.)$, and prediction-head $p(.)$ and evaluate in downstream tasks in 2 different settings similar to \cite{saeed2020contrastive} where 1) We use DECAR embeddings as a feature extractor and only train a linear classifier with the frozen embeddings 2) We fine-tune the entire End-to-End network on the downstream task. Importantly, we assess the performance on several diverse
datasets to determine the transferability of learned representations across audio domains and recording conditions. We show our results in \ref{sec:results}.

We train all our downstream tasks with a similar hyper-parameter setting, where we use Adam optimizer, a learning rate of $10^{-3}$, batch size of 64, and train it for a maximum of 50 epochs.

\section{Results}
\label{sec:results}

Table \ref{tab:test_dev_results} reports accuracy of \textbf{DECAR} embeddings over 9 downstream classification tasks, with \textbf{PIC} as our clustering algorithm in self-supervised pre-training. As discussed earlier, we report results both on frozen and fine-tuned DECAR embeddings for each downstream task and repeat the same for randomly initialized weights, which act as our baseline. As we can see, a linear classifier trained over frozen DECAR embeddings outperforms drastically the same setting when trained with randomly initialized embeddings. This makes it evident that we learn powerful embeddings from our pre-training step. Next, we also fine-tune our pre-trained DECAR embeddings on each downstream task and show significant gains compared to training with randomly initialized weights. We show a boost in performance in both speech and non-speech tasks showing that our DECAR pre-training learns representations that generalize well across a wide range of audio tasks. Our results show better improvements in low-resource downstream settings like IEMOCAP, Voxceleb, and Acoustic Scenes with the limited number of samples per class when compared to tasks like Speech Commands and BSD. Additionally, we observe better improvements in speech tasks where the prosodic features in speech aid the model to learn to distinguish among different classes, like SER, versus the ones which require sequence information like Speech Commands.

Our results are not directly comparable with the prior art in this domain due to 2 main reasons:
\begin{enumerate}
    \item  There is no standard benchmark for comparing general-purpose audio representations and the presence of several nuances around dataset settings in certain tasks.
    \item We pre-train DECAR in a low-resource setting compared to most other work in self-supervised audio and speech representation learning. We have pre-trained using only 0.2 million (10$\%$) data samples compared to 2 million used in \cite{saeed2020contrastive}. We acknowledge the fact that more data can further boost the performance in all tasks, including high-resource downstream tasks.
\end{enumerate}



\begin{table}[ht]
\small
\centering
\footnotesize
\caption{Test accuracy (\%) on downstream tasks}
\begin{tabular}{l@{ }c@{ }c@{ }c@{ }c}
\hline
 & \multicolumn{2}{c}{\textbf{Random Init.}} &\multicolumn{2}{c}{\textbf{DECAR}} \\
\textbf{Task} & \textbf{Frozen} & \textbf{Fine-tuned} & \textbf{Frozen} & \textbf{Fine-tuned} \\
\hline
Speaker Id. (LBS) & 4.2 & 94.6 & 62.5 & \textbf{97.0} \\
Speaker Id. (Voxceleb) & 0.2 & \textbf{57.6} & 2.5 & 57.5 \\
Speech Commands (V1) & 62.0 & 97.4 & 63.9 & \textbf{97.6} \\
Speech Commands (V2) & 62.9 & \textbf{97.7} &  65.7 & 97.6 \\
Acoustic Scenes & 32.6 & 61.6 & 51.4 &  \textbf{66.9}\\
Speech Emotion & 52.3 & 61.8 & 60.5  &  \textbf{62.4} \\ 
Birdsong Detection & 67.6 & 90.0 & 76.4 & \textbf{90.3} \\
Language ID & 26.3 & 73.4 & 46.0 & \textbf{76.5} \\
Music Instrument & 23.6 & \textbf{78.6} & 59.9 & 78.4 \\
\hline
\end{tabular}
\label{tab:test_dev_results}
\end{table}

\section{Ablation Study}
\label{sec:results-analysis}

Clustering being an optimal part of our self-supervised learning algorithm, we wanted to study how variations in the clustering algorithm affected performance on downstream tasks. For this, we compare the performance of our downstream tasks across PIC and K-Means clustering algorithms. Table \ref{tab:test_dev_results_cluster} shows that though PIC dominates in most cases, there is a negligible difference in performance when it comes to K-Means. This shows that our system is robust to the clustering algorithm used. Additionally, we also study the effect of the hyperparameter $k$, or the number of centroids, on our downstream performance for K-Means. We notice no particular trend here and would want to address this as part of our future work with the entire Audioset.

\begin{table}[ht]
\small
\centering
\caption{Impact of cluster variations on test accuracy(\%)}
\begin{tabular}{l l c c c}
\hline
 & \multicolumn{3}{c}{\textbf{K-Means}} & \textbf{PIC}\\
\textbf{Task} & \textbf{256} & \textbf{512} & \textbf{1024}  &\\
\hline
Speaker Id. (LBS) & 96.6 & 96.2 & 97.3 & \textbf{97.0}\\
Speaker Id. (Voxceleb) & 53.4 & 49.7 & 57.0 & \textbf{57.5}\\
Speech Commands (V1) & 97.3 & 97.0 & 97.4 & \textbf{97.6} \\
Speech Commands (V2) & 97.4 & 97.3 & 97.4 & \textbf{97.6} \\
Acoustic Scenes & 68.2 & \textbf{68.7} & 66.5 & 66.9\\
Speech Emotion & 62.1 & \textbf{64.0} & 63.1 & 62.4\\
Birdsong Detection & 89.6 & 89.5 & 90.1 & \textbf{90.3}\\
Language ID & 75.4 & 75.0 & 73.9 & \textbf{76.5}\\
Music Instrument & 77.3 & 77.5 & 78.1 & \textbf{78.4}\\
\hline
\end{tabular}
    
\label{tab:test_dev_results_cluster}
\end{table}

\section{Conclusion and Future Work}
\label{sec:conclusion}

In this paper, we present DECAR, a simple and easy-to-implement self-supervised audio representation learning algorithm that alleviates the need for large batches or any additional steps and prior domain knowledge. Our system shows significant performance boosts in low resource downstream settings even when pre-trained on a fraction of the total amount of data used in prior work. Future work would include mitigating the need for an additional forward pass by learning cluster assignments online, thus allowing it to scale to practically unlimited amounts of data, applying better augmentation schemes during the cluster assignment prediction step, and also increase the diversity of downstream tasks.

\clearpage



\bibliographystyle{IEEEbib}
\bibliography{strings,refs}

\end{document}